%% file: jem.tex
\documentclass[graybox]{svmult}

\usepackage{mathptmx}       
\usepackage{helvet}         
\usepackage{courier}        
\usepackage{type1cm}        
\usepackage{makeidx}         
\usepackage{graphicx}        
\usepackage{multicol}        
\usepackage[bottom]{footmisc}

\usepackage{amssymb, amsmath, latexsym, amsfonts, mathcomp}

\usepackage[style=numeric,url=false,doi=false,isbn=false,maxnames=10,maxbibnames=10]{biblatex}
\bibliography{Everything}
\makeindex             

\usepackage{dcolumn}
\usepackage{caption}
\usepackage{microtype}

\input{macros}

\let\oldmarginpar\marginpar
\renewcommand\marginpar[1]{\-\oldmarginpar[\raggedleft\tiny #1]{\raggedright\footnotesize #1}}

\begin{document}

\title*{The Geometry of Radiative Transfer}
\author{Christian Lessig and Alex L. Castro}
\institute{Christian Lessig \at Department of Computing$\,+\,$Mathematical Sciences, California Institute of Technology, now with the University of Toronto and Technische Universit{\"a}t Berlin,  \email{lessig@caltech.edu};
\and Alex L. Castro \at Departamento de Matem{\'a}tica, Pontif{\'i}cia Universidade Cat{\'o}lica do Rio de Janeiro, \email{alex.castro@mat.puc-rio.br}}
%
%
\maketitle

\abstract{
We present the geometry and symmetries of radiative transfer theory. Our geometrization exploits recent work in the literature that enables to obtain the Hamiltonian formulation of radiative transfer as the semiclassical limit of a phase space representation of electromagnetic theory. Cosphere bundle reduction yields the traditional description over the space of "positions and directions", and geometrical optics arises as a special case when energy is disregarded. It is also shown that, in idealized environments, radiative transfer is a Lie-Poisson system with the group of canonical transformations as configuration space and symmetry group.
}



\input{intro}

\input{derivation}

\input{connections}

\input{conclusion}

\begin{acknowledgement}
We thank Eugene Fiume, Mathieu Desbrun, Tudor Ratiu, Boris Khesin, and Jerry Marsden for discussions. 
C.L acknowledges support by the National Science and Engineering Council of Canada, as well as MITACS and GRAND National Centres of Excellence. C.L. was also supported by NSF grant CCF-1011944.
A.C. thanks the University of Toronto and the Fields Institute for their hospitality and a postdoctoral fellowship during which this work was initiated. 
\end{acknowledgement}

\printbibliography

\end{document}

%% file: macros.tex
\def\Den{\mathrm{Den}}
\def\E{\mathbf{E}}

\def\R{\mathbb{R}}

\def\g{\mn{g}}                

\def\eps{\epsilon}

\def\vdiv{\mathrm{div}}

\DeclareMathOperator*{\tr}{tr}

\def\Dgc{\mathrm{Diff}_{\mathrm{can}}}

\def\Dgv{\mathrm{Diff}_{\mu}}

\newcommand{\SO}[1]{\ensuremath{\mathrm{SO}(#1)}}

\def\mmap{\ensuremath \mathbf{J}}

\def\Lie{\pounds}

\def\ad{\mathrm{ad}}

\def\g{\mathfrak{g}}



%% file: intro.tex
\section{Introduction}

\begin{figure*}[t]
  \setlength{\abovecaptionskip}{-10pt}
  \setlength{\belowcaptionskip}{-5pt}
  \begin{center}
  \centerline{
  \includegraphics[trim = 0mm 0mm 0mm 0mm, clip, scale=0.7]{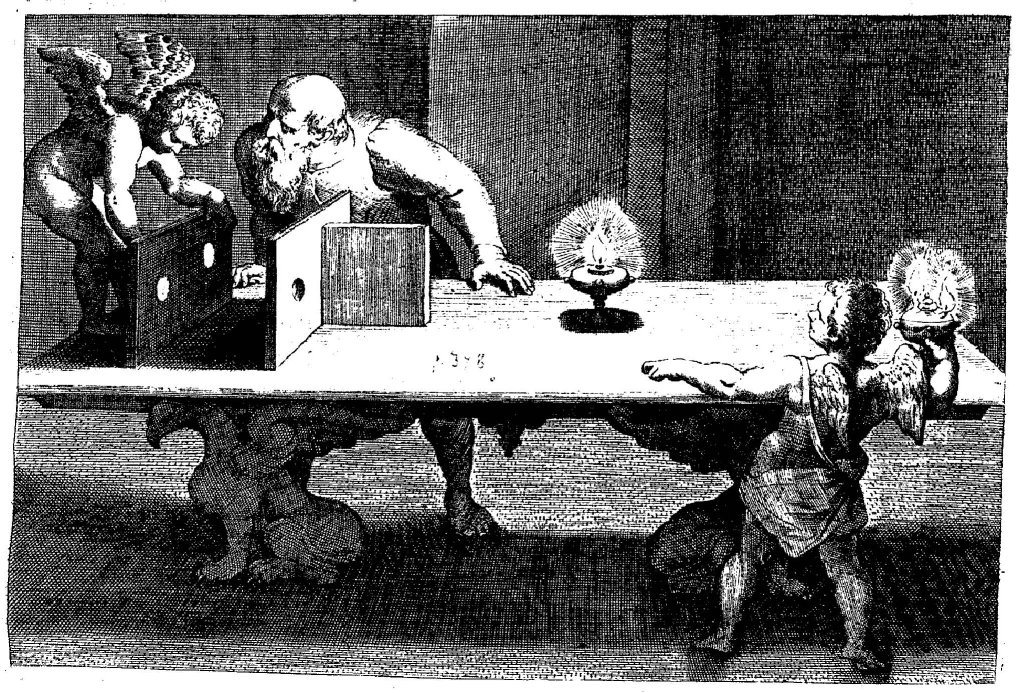}}
  \end{center}
  \caption{Early research in radiometry as illustrated by Rubens (from~\cite{Lambert2001}).}
  \label{fig:rubens}
\end{figure*}

Radiative transfer describes the transport of electromagnetic energy in macroscopic environments, classically when polarization effects are neglected~\cite{Pomraning1973}.
The theory originates in work by Bouguer~\cite{Bouguer1729,Bouguer1760} and Lambert~\cite{Lambert2001} in the 18$^{\textrm{th}}$ century where light intensity and its measurement were first studied systematically, cf. Fig.~\ref{fig:rubens}. 
In the 19$^\textrm{th}$ and early 20$^\textrm{th}$ century the theory was then extended to include transport and scattering effects~\cite{Lommel1889,Chwolson1889,Schuster1905,Schwarzschild1906}.
To this day, however, radiative transfer is a phenomenological theory with a mathematical formulation that still employs the concepts introduced by Lambert in the 18$^{\textrm{th}}$ century---and this despite the importance of the theory in a multitude of fields, such as medical imaging, remote sensing, computer graphics, atmospheric science, climate modelling, and astrophysics.

\begin{figure*}[t]
  \setlength{\abovecaptionskip}{-10pt}
  \setlength{\belowcaptionskip}{-5pt}
  \begin{center}
    \centerline{\hspace{-0.05in}
    \includegraphics[trim = 50mm 330mm 30mm 20mm, clip, scale=0.36]{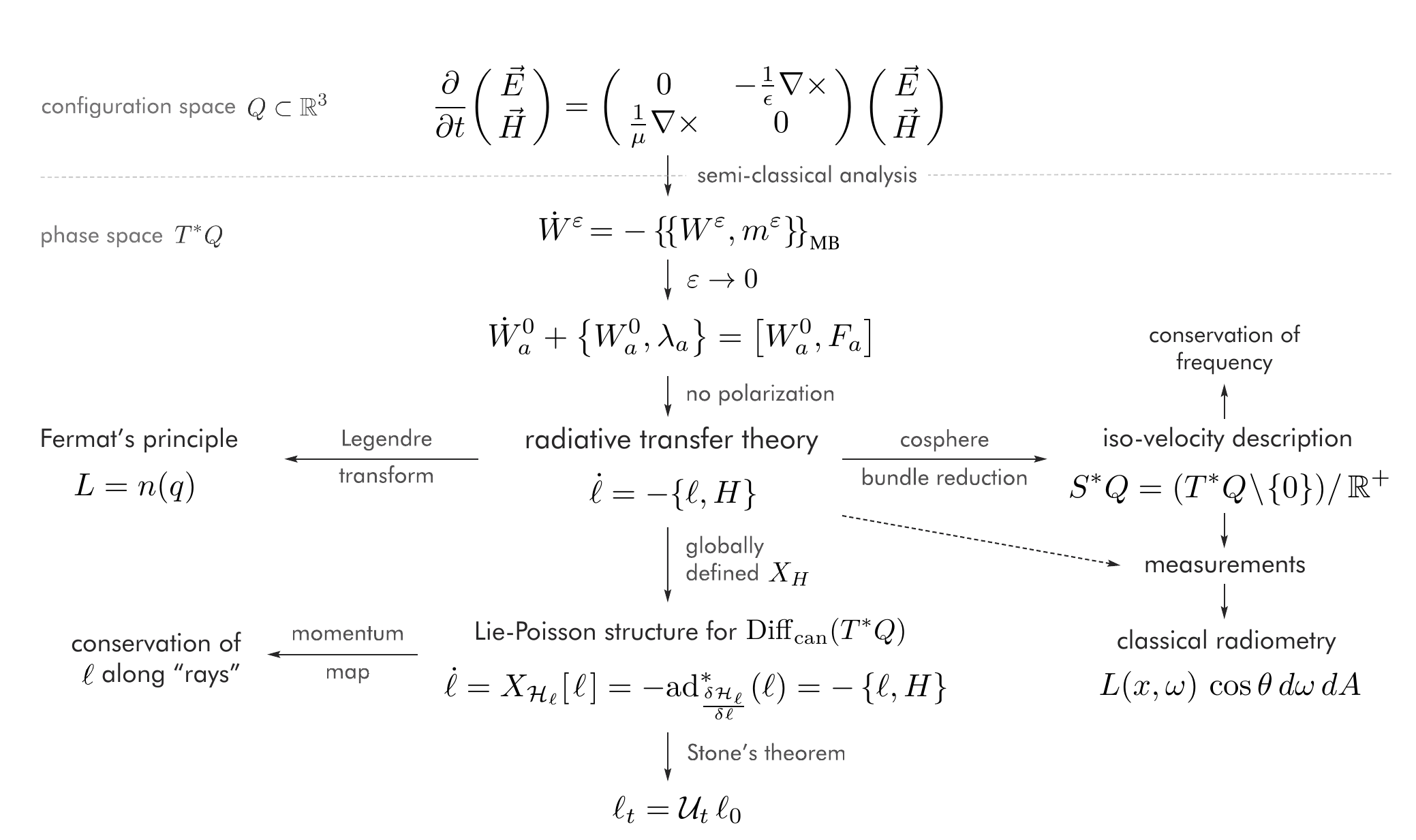}}
  \end{center}
  \caption{Overview of the physical foundations and the geometric structure of radiative transfer theory. Semi-classical analysis yields a description of Maxwell's equations on phase space $T^*Q$ where the electromagnetic field $\vec{F} = (\vec{E},\vec{H})^T$ is represented by the Wigner transform $W^{\varepsilon}$ and dynamics are governed by a matrix-valued analogue of the Moyal bracket $\{ \! \{ \, , \} \! \}_{\textrm{MB}}$. When the short wavelength limit is considered, this leads to a transport equation for polarized light with the $2 \times 2$ matrix density $W_a^0$ being formed by the classical Stokes parameters.
  When polarization is also neglected, $W_a^0$ becomes the scalar light energy density $\ell \in \Den(T^*Q)$ whose dynamics are governed by the Poisson bracket. The classical five dimensional formulation of radiative transfer is obtained using cosphere bundle reduction with the light frequency being the associated conserved quantity. When radiative transfer is considered globally and the light energy density $\ell_t$ at time $t$ forms a configuration of the system, radiative transfer becomes a Lie-Poisson system for the group $\Dgc(T^*Q)$ of canonical transformations.}
  \label{fig:overview}
\end{figure*}

In the following, we will explain the physical foundations of radiative transfer in media with varying refractive index and we study the geometry of the theory and its symmetries; an overview is provided in Fig.~\ref{fig:overview}.
Following recent advances in applied mathematics, semi-classical analysis is employed to lift classical electromagnetic theory from configuration space $Q \subseteq \R^3$ to phase space $T^*Q$.
By restricting the dynamics on $T^*Q$ to a non-zero energy level and considering the short wavelength limit one obtains a transport equation for polarized light, and further only considering the energy that is transported and neglecting polarization leads to radiative transfer theory in a Hamiltonian formulation.
Our derivation shows that the central quantity of radiative transfer theory is the phase space light energy density $\ell \in \Den(T^*Q)$ and that radiance, which plays this role in the classical formulation, is meaningful only in the context of measurements, the setting Lambert considered when he introduced the concept~\cite{Lambert2001}.
With the Hamiltonian formulation of radiative transfer on $6$-dimensional phase space $T^*Q \cong \R^3 \times U$, the classical $5$-dimensional description over the space of ``positions and directions'' is obtained when the conservation of frequency is exploited.
The associated symmetry enables the reduction of the dynamics from the cotangent bundle $T^*Q$ to the cosphere bundle $S^*Q = (T^*Q \! \setminus \! \{ 0 \}) / \R^+$ and time evolution is then described by contact dynamics.
Fermat's principle, the Lagrangian formulation of geometrical optics, is obtained from the Hamiltonian formulation of radiative transfer through a non-canonical Legendre transform when energy transport is neglected.
From our point of view, geometric optics is thus a special case of radiative transfer theory.
When radiative transfer is considered from a global perspective with the light energy density $\ell_t \in \Den(T^*Q)$ at time $t$ as a configuration of the system, the configuration space of the theory becomes the group $\Dgc(T^*Q)$ of canonical transformation. 
Radiative transfer has then a Lie-Poisson structure and the associated symmetry is the conservation of light energy density along trajectories in phase space.
This provides a modern rationale for the classical law of ``conservation of radiance along a ray''~\cite{Nicodemus1963}.
It also reveals a surprising similarity to Kelvin's circulation theorem in ideal fluid dynamics.

%% file: derivation.tex
\section{A Modern Formulation of Radiative Transfer Theory}
\label{sec:derivation}

Following recent work in the literature, in this section we will describe how the Hamiltonian formulation of radiative transfer arises as the asymptotic limit of Maxwell's equations and we will also study the geometry and symmetries of the system.

\subsection{Derivation}

From the scale hierarchy of electromagnetic radiation in physics it is apparent that radiative transfer has to arise at the short wavelength limit of Maxwell's equations, Hamilton's equations for electromagnetic field theory~\cite[p. 24]{Ratiu1999}.
Nonetheless, the exact correspondence was open for more than 200 years and still in the 1990s Mandel and Wolf~\cite{Mandel1995} lamented that ``in spite of the extensive use of the theory of radiative energy transfer, no satisfactory derivation of its basic equation from electromagnetic theory has been obtained up to now''.\footnote{The first derivation of geometric optics from Maxwell's equations goes back to Sommerfeld and Runge~\cite{Sommerfeld1911}, cf.~\cite[Chapter III]{Born1985} for historical details. Derivations of geometrical optics do not provide long time transport equations for the light energy density, cf. also~\cite{Tartar2010}.}
Recent work in applied mathematics~\cite{Tartar1990a,Tartar1990b,Gerard1991,Gerard1997,Ryzhik1996} fills this gap and in the following we will summarize a rigorous derivation of radiative transfer theory from Maxwell's equations.

In a source free region $Q \subset \R^3$, Maxwell's equations are given by~\cite{Born1985}
\begin{subequations}
\label{eq:maxwell:explicit}
\begin{align}
  \dfrac{\partial}{\partial t} \! \left( \! \! \begin{array}{c} \vec{E} \\ \vec{H} \end{array} \! \! \right) 
  =&
  \left( \! \! \begin{array}{cc} 0 & -\frac{1}{\epsilon} \nabla \times \\ \frac{1}{\mu} \nabla \times & 0 \end{array} \! \! \right) \left( \! \! \begin{array}{c} \vec{E} \\ \vec{H} \end{array} \!  \! \right)
    \\[7pt]
    \vdiv{(\vec{E})} &= 0 \quad \quad \vdiv{(\vec{H})} = 0 
\end{align}
\end{subequations}
where $\eps : Q \to \R$ and $\mu : Q \to \R$ are the electric permittivity and magnetic permeability, respectively, and $\vec{E}$ and $\vec{H}$ represent the electric and magnetic fields; in the following it will be understood that these fields are divergence free.
By introducing $\vec{F} = (\vec{E},\vec{H})^T$, Eq.~\ref{eq:maxwell:explicit} can be written as
\begin{align}
  \dot{\vec{F}} = \vec{M} \, \vec{F}
  \label{eq:maxwell:operator}
\end{align}
and we will denote $\vec{M}$ as the Maxwell operator.\footnote{Eq.~\ref{eq:maxwell:operator} is closely related to the spacetime formulation of electromagnetic theory with $\vec{F}$ being the components of the Faraday $2$-form $F = E \wedge dt + B$, cf.~\cite[Sec. 3]{Gay-Balmaz2008}.}
The classical observable of electromagnetic theory is the energy density $\mathcal{E}(q,t)$, given by
\begin{align}
  \mathcal{E}(q,t) = \Vert \vec{F} \Vert_{\varepsilon,\mu}^2 = \frac{\varepsilon}{2} \Vert \vec{E} \Vert^2 + \frac{\mu}{2} \Vert \vec{H} \Vert^2 .
  \label{eq:em_energy_density}
\end{align}
We are interested in the transport of the energy density $\mathcal{E}(q,t)$ in macroscopic environments. 
To describe this regime mathematically we introduce the scale parameter \begin{align}
  \varepsilon = \lambda / d_n .
  \label{eq:def:small_parameter}
\end{align}
In Eq.~\ref{eq:def:small_parameter}, $\lambda$ is the wavelength of light and $d_n$ the average distance over which the refractive index $n = \sqrt{\eps \, \mu} : Q \to \R$ varies, cf.~\cite[Chapter 22.5]{Misner1973}.
In macroscopic environments one has $\lambda \ll d_n$ and asymptotically these can thus be studied by letting $\varepsilon \to 0$.
In the following, we will often write $\vec{F}^{\varepsilon}$, $\mathcal{E}^{\varepsilon}$  etc. to make the dependence of variables on the scale parameter explicit.

The classical approach to study short wavelength asymptotics is the Wenzel-Kramers-Brillouin (WKB) approximation.\footnote{It is by now well known that the WKB approximation goes back to work by Liouville and Green in the first half of the 19$^\textrm{th}$ century.}
However, this ansatz is limited in that solutions are only well defined until caustics form, at which point the approximation becomes multi-valued, and that the initial conditions must satisfy the WKB form $u^{\varepsilon}(q,t) = a(q,t) \, e^{i S(q,t) / \varepsilon}$.
Additionally, Maxwell's equations describe the time evolution of the field $\vec{F}$, while we are interested in the limit $\varepsilon \to 0$ of the energy density $\mathcal{E}^{\varepsilon}(q,t)$ that depends quadratically on the field.
This provides a serious obstruction for any approach to determine the transport of the limit energy density $\mathcal{E}^0(q,t)$~\cite{Sparber2003}.

The limitations of classical approaches to describe the transport of the limit energy density $\mathcal{E}^{0}(q,t)$ can be circumvented by lifting electromagnetic theory to phase space $T^*Q$ and studying the short wavelength limit there~\cite{Gerard1997,Ryzhik1996}.
The electromagnetic field $\vec{F}$ can be lifted to $T^*Q$ using the Wigner transform~\cite{Wigner1932,Gerard1997}, yielding a $6 \times 6$ matrix density $W^{\varepsilon}[\vec{F}]$ whose components are given by
\begin{align}
  W^{\varepsilon}[\vec{F}]_{ij}^{\varepsilon}(q,p) = 
  \frac{1}{(2 \pi)^3} \int_Q e^{i p \cdot r} \, \vec{F}_i^{\varepsilon}(q - \frac{\varepsilon}{2} r) \, \vec{F}_j^{\varepsilon}(q + \frac{\varepsilon}{2} r) \, dr .
\end{align}
The lift of the Maxwell operator $\vec{M}^{\varepsilon}$ to phase space is provided by its matrix-valued symbol $m^{\varepsilon}(q,p)$ which can formally be expanded as  $m^{\varepsilon}(q,p) = m^0(q,p) + \eps m^1(q,p) + \eps^2 m^2(q,p) + \cdots$.
Time evolution on phase space is described by 
\begin{align}
  \dot{W}^{\varepsilon} = - \{ \! \{ W^{\varepsilon} , m^{\varepsilon} \} \! \}_{\textrm{MB}}
\end{align}
where $W^{\varepsilon} \equiv W^{\varepsilon}[\vec{F}]$ and $\{ \! \{ \, , \} \! \}_{\textrm{MB}}$ is a matrix-valued ``Moyal bracket''~\cite[Eq. 6.12]{Gerard1997}.
Expanding this bracket one obtains
\begin{align}
  \dot{W}^{\varepsilon} = \frac{1}{\eps} [W^{\varepsilon} , m^{\varepsilon}] - \frac{1}{2 i}\left( \left\{ W^{\epsilon} , m^{\varepsilon} \right\} - \left\{ m^{\varepsilon} , W^{\epsilon} \right\} \right) + \mathcal{O}(\epsilon)
  \label{eq:moyal_bracket:expanded}
\end{align}
where $\{ \, , \}$ is commonly denoted as a matrix-valued ``Poisson bracket'';\footnote{The matrix-valued ``Poisson bracket'' is computed by performing matrix multiplication with scalar multiplication replaced by the usual Poisson bracket, see for example~\cite[Appendix A]{Teufel2003}.} it is not a Poisson bracket in the formal sense and we will return to this point in Sec.~\ref{sec:conclusion}.
In contrast to the scalar Moyal bracket where the commutator in the first term vanishes by the commutativity of multiplication in the algebra $\mathcal{F}(T^*Q)$, in the matrix-valued case care is needed that the first term does not diverge as $\varepsilon \to 0$.
This divergence can be circumvented by restricting dynamics to the eigenspaces of the Maxwell symbol $m^{\varepsilon}$.
The diagonally identical symbol matrix is then in the ideal of the matrix algebra and the appropriately restricted commutator $[W^{\varepsilon},m^{\varepsilon}]$ hence vanishes; this is the matrix-valued analogue of the Bohr-Sommerfeld quantization condition~\cite{Littlejohn1991a}.
The eigenvalues of the Maxwell symbol are given by~\cite{Ryzhik1996}
\begin{align}
  \lambda_0 = 0 
  \quad\quad\quad \lambda_1 = \frac{c}{n(q)} \Vert p \Vert 
  \quad\quad\quad \lambda_2 = - \frac{c}{n(q)} \Vert p \Vert ,
\end{align}
each having multiplicity two. 
Only $\lambda_1$ and $\lambda_2$ have physical significance, corresponding to forward and backward propagation in time.
We will denote the projection onto the eigenspace associated with $\lambda_a$ by $\Pi_a$, with $a \in \{1,2\}$.
Projecting the Wigner distribution $W^{\varepsilon}$ onto the $a^{\textrm{th}}$ eigenspace and taking the limit $\varepsilon \to 0$ yields
\begin{align}
  W_a^{0} = \Pi_a W^0 \Pi_a 
  = \frac{1}{2}
  \left[
  \begin{array}{cc}
    I + Q & U + iV \\
    U - iV & I - Q
  \end{array}
  \right] dq \, dp .
  \label{eq:wigner_projected}
\end{align}
The parameters $I,Q,U,V$ in Eq.~\ref{eq:wigner_projected} are the Stokes parameters for polarized light.
This provides much physical intuition for the projected limit Wigner distribution $W_a^0$. 
From Eq.~\ref{eq:moyal_bracket:expanded} one obtains for the time evolution of $W_a^0$ that~\cite{Gerard1997}
\begin{align}
  \dot{W}_a^{0} 
  = \Pi_a \left\{  W_a^{0} , \lambda_a \right\} \Pi_a + \left[ W_a^0 , \Pi_a m^1 \Pi_a \right]
  = \left\{  W_a^{0} , \lambda_a \right\} + \left[ W_a^0 , F_a^0 \right]
  \label{eq:transport:polarization}
\end{align}
where $F_a^0 = [ \Pi_a , \{ \lambda_a , \Pi_a \}] + \Pi_a m^1 \Pi_a$ and, as before, $m^1$ is the first order term in the formal expansion of the symbol $m^{\varepsilon}$ in the order parameter $\epsilon$.
Intuitively, the ``Poisson bracket'' $\left\{  W_a^{0} , \lambda_a \right\}$ describes the transport of the polarized radiation $W_a^0$ on phase space while the commutator $\left[ W_a^0 , F_a^0 \right]$ is responsible for the change in polarization during transport, we will again come back to this in Sec.~\ref{sec:conclusion}.
Classical radiative transfer is a scalar theory and does not consider polarization. 
For unpolarized light the Stokes parameters satisfy $Q,U,V = 0$.
The matrix density $W_a^0$ is then completely described by its trace, representing the intensity $I$ of the radiation.
We thus define the \emph{light energy density} as
\begin{align}
  \ell = \tr{(W_i^0)} = \mathcal{L}(q,p) \, dq \, dp \in \Den(T^*Q) .
\end{align}
It follows from Eq.~\ref{eq:transport:polarization} that the transport of $\ell \in \Den(T^*Q)$ is described by 
\begin{align}
  \dot{\ell} = -\left\{ \ell , H \right\} 
  \label{eq:lte}
\end{align}
with the Hamiltonian $H \in \mathcal{F}(T^*Q)$ being the eigenvalue $\lambda_a$, that is
\begin{align}
  H(q,p) = \pm \frac{c}{n(q)} \Vert p \Vert .
  \label{eq:hamiltonian}
\end{align}
The light energy density $\ell \in \Den(T^*Q)$ is related to the limit electromagnetic energy density $\mathcal{E}(q) \in \Den(Q)$ by the fiber integral
\begin{align}
  \lim_{\eps \to 0}\mathcal{E}^{\eps}(q,t) 
  = \int_{T_q^*Q} \ell
  = \int_{T_q^*Q} \mathcal{L}(q,p,t) \, dp 
  \label{eq:energy_density:fiber_integral}
\end{align}
and $\ell \in \Den(T^*Q)$ can be understood as an angularly resolved form of the electromagnetic energy density.
Hence, the light energy density $\ell \in \Den(T^*Q)$ together with Eq.~\ref{eq:lte} provide the sought after system to describe the transport of the limit energy density $\mathcal{E}^0(q,t)$.  
Eq.~\ref{eq:lte} describes the transport of electromagnetic energy in macroscopic environments to good approximation, as is evidenced by the success of radiative transfer in a wide range of fields.

\subsection{Cosphere Bundle Reduction for Radiative Transfer}

Eq.~\ref{eq:lte} describes radiative transfer theory as a Hamiltonian system on $6$-dimensional phase space $T^*Q \cong \R^3 \times Q$. 
In the literature, however, the theory is usually defined over the $5$-dimensional space of ``positions and directions''.
The two descriptions are related through the symmetry associated with the well known conservation of frequency during transport.
The Hamiltonian in Eq.~\ref{eq:hamiltonian} is homogeneous of degree one in the momentum, $H(q, \alpha p) = \alpha H(q,p)$, for $\alpha \in \R^+$, and, moreover, momentum and light frequency are proportional. 
This suggests that the symmetry group associated with the conservation of frequency is $(\R^+, \cdot)$ acting on the fibers $T_q^* Q$ by $m_{\alpha}(q,p) = (q,\alpha \, p)$.
As is well known~\cite[App. 4]{Arnold1989}, the quotient space for this action is given by the cosphere bundle
\begin{align}
  S^*Q = \left( T^*Q \! \setminus \! \{ 0 \} \right) / \R^+  
\end{align}
and for a Hamiltonian of degree one dynamics on $T^*Q$ drop to a contact Hamiltonian flow along the Reeb vector field on $S^*Q$~\cite{Ratiu1981}.
The cosphere bundle $S^*Q$, identified with the sphere bundle $S^2 Q$ using the standard metric in $\R^3$, provides a modern interpretation for the classical space of ``positions and directions'', and the homogenity of the Hamiltonian explains why such a description on $S^*Q$ is possible, despite the Hamiltonian character of the system that seemingly requires a description on an even dimensional space.

\subsection{Radiative Transfer as a Lie-Poisson System}

\begin{figure}[t]
\setlength{\abovecaptionskip}{-10pt}
\setlength{\belowcaptionskip}{-10pt}
  \begin{center}
    \centerline{\hspace{-0.1in}
    \includegraphics[trim = 30mm 195mm 65mm 35mm, clip, scale=0.9]{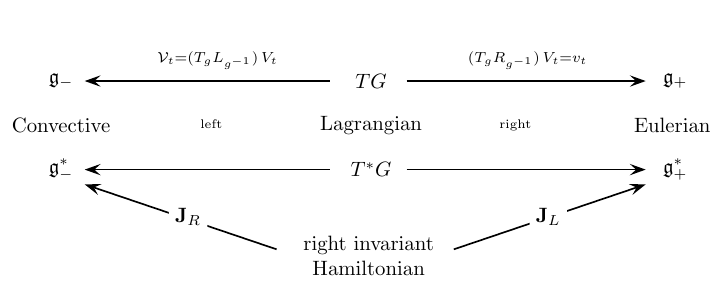}}
  \end{center}
  \caption{The structure of Lie-Poisson systems. Classical examples of such systems are the rigid body, where the Lie group is $\SO{3}$ with a left invariant Hamiltonian, and the ideal Euler fluid, where the Lie-group is the group $\Dgv(Q)$ of volume preserving diffeomorphisms with a right invariant Hamiltonian~\cite{Arnold1966,Ebin1970}.}
  \label{fig:lie_poisson}
\end{figure}

A central result in classical radiative transfer theory is the ``conservation of radiance along a ray''~\cite{Goodman2010}.
The symmetry associated with this conservation law becomes apparent when the light energy density $\ell_t$ at time $t$, globally over all $T^*Q$, is considered as one configuration of the system.
Time evolution can then be described by the pullback $\ell_t = \eta_t^* \ell_0$ along the map $\eta_t : T^*Q \to T^*Q$ that is generated by the flow of the Hamiltonian vector field $X_H$ defined by Eq.~\ref{eq:hamiltonian}, and when $X_H$ is defined globally the set of all such maps $\eta_t$ forms the infinite dimensional Lie group $\Dgc{(T^*Q)}$ of canonical transformations~\cite{Ebin1970}.
With respect to the initial light energy density $\ell_0$ all physically valid configurations $\ell_t$ can then be described by an element $\eta_t$ in $\Dgc(T^*Q)$ and the group becomes the configuration space of radiative transfer.
We thus have 
Radiative transfer is then a Lie-Poisson system for $\Dgc(T^*Q)$, cf. Fig.~\ref{fig:lie_poisson}.

The Lie-Poisson structure for the group $\Dgc(T^*Q)$ was first studied by Marsden and coworkers in the context of plasma physics~\cite{Marsden1982a}. 
The Lie algebra $\g$ of $\Dgc(T^*Q)$ are infinitesimal canonical transformations, that is $\g \cong \mathfrak{X}_{\textrm{\tiny{Ham}}}(T^*Q)$, and by identifying the Hamiltonian vector fields with the generating Hamiltonian functions, $\g \cong \mathcal{F}(T^*Q)$, the dual Lie algebra $\g^*$ becomes $\Den(T^*Q)$.\footnote{We disregard here some technical details in the construction of the dual Lie algebra. See for example~\cite[Chapter 2.3.5.3]{Lessig_phd}.}
With $\g^* \cong \Den(T^*Q)$, it is natural to consider the light energy density $\ell$ as an element in $\g^*$.
The time evolution of $\ell$ is then described by coadjoint action $\Dgc(T^*Q) \times \g_+^* \to \g_+^*$ in the Eulerian representation and infinitesimally this is given by~\cite[Sec. 3.3]{Lessig_phd}
\begin{align}
  \dot{\ell}
  = \ad_{\frac{\delta \mathcal{H}}{\delta \ell}}^* \ell 
  = \ad_{H}^* \ell 
  = -\left\{ \ell , H \right\} ;
  \label{eq:infin_coadjoint}
\end{align}
indeed, that the Poisson bracket describes infinitesimal coadjoint action $\ad^* : \g \times \g^* \to \g^*$ for $\Dgc(T^*Q)$ is an a posterior justification for considering the group as the configuration space for ideal radiative transfer~\cite[Sec. 6]{Marsden1982a}.
In Eq.~\ref{eq:infin_coadjoint}, $\mathcal{H} \equiv \mathcal{H}[\ell]$ is the field Hamiltonian 
\begin{align}
  \mathcal{H}[\ell] = \int_{T^*Q} \ell(q,p) \, H(q,p) \, dq \, dp
\end{align}
which is the density weighted integral of the ``single particle'' Hamiltonian $H(q,p)$ in Eq.~\ref{eq:hamiltonian}.
With the light energy density as an element in the dual Lie algebra, it follows immediately from the general theory of Lie-Poisson systems that the momentum map $\mmap_R$ is the convective light energy density and that this quantity is conserved~\cite[Theorem 11.4.1]{Ratiu1999}, cf. Fig.~\ref{fig:lie_poisson}.\footnote{A direct proof can be found in~\cite[Chapter 3.3]{Lessig_phd}.}
By the change of variables theorem, this can be interpreted as conservation of light energy density along trajectories in phase space and it provides a modern formulation and justification for ``conservation of radiance along a ray'' in the classical literature.
Interestingly, with the Lie-Poisson structure a close formal analogy between ideal radiative transfer and ideal fluid dynamics exists, cf. Table~\ref{tab:comp:fluid_light}.

\newcolumntype{C}{>{\centering\arraybackslash} m{3.5cm} }
\begin{table}[t]
\setlength{\abovecaptionskip}{5pt}
\setlength{\belowcaptionskip}{-10pt}
\centering
\begin{tabular*}{1.0\textwidth}{>{\centering\arraybackslash}m{4.0cm} |C|C m{0.01in}}
  & fluid dynamics & radiative transfer &
  \\[0.08in]
  \hline
  Lie group & $\Dgv(Q)$ & $\Dgc(T^*Q)$ &
  \\[0.08in]
  \hline
  Lie algebra & 
  $\mathfrak{X}_{\mathrm{\tiny{div}}}(Q)$ &  $\mathfrak{X}_{\mathrm{\tiny{Ham}}}(T^*Q)$ &
  \\[0.08in]
  \hline
  dual Lie algebra & $\omega \in \Omega^2(Q)$ & $\ell \in \Den(T^*Q)$ &
  \\[0.1in]
  \hline
  coadjoint action & $\dot{\omega} = - \Lie_v \omega$ & $\dot{\ell} = -\Lie_{X_H} \ell$ &
  \\[0.08in]
  \hline
  classical conservation law & Kelvin's theorem & conservation of radiance &
  \\[0.08in]
\end{tabular*}
\caption{Correspondence between ideal fluid dynamics and ideal radiative transfer. The fluid velocity is denoted by $v \in \mathfrak{X}_{\textrm{div}}(Q)$ and $\omega \in \Omega^2(Q)$ is the fluid vorticity.}
\label{tab:comp:fluid_light}
\end{table}

Next to the transport on $T^*Q$, the time evolution of radiative transfer can also be understood as a functional analytic flow on the space of light energy densities.
By identifying the Hamiltonian vector field $X_H$ with an anti-self-adjoint operator, Stone's theorem~\cite[Theorem 6.2.18.3]{Marsden1983} enables us to describe radiative transfer as
\begin{align}
  \ell_t = \eta_t^* \ell_0 = U_t \, \ell_0
  \label{eq:functional_analytic_flow}
\end{align}
where $U_t$ is a unitary operator.
Such a functional analytic representation of the action of an infinite dimensional diffeomorphism group is often referred to as Koopmanism~\cite{Koopman1931}, cf. also~\cite[Chapter 8.4]{Marsden2004}.
An interesting aspect of Eq.~\ref{eq:functional_analytic_flow} is that it provides a rigorous basis for the operator formulation of radiative transfer that can be found in the classical literature, see for example~\cite{Duderstadt1979}.
Eq.~\ref{eq:functional_analytic_flow} also provides a natural starting point to include scattering effects, for example at surfaces, that do not have a geometric but a well known functional analytic description.

%% file: connections.tex
\section{Some Connections to the Classical Formulations}

In this section, we will relate our geometric formulation of radiative transfer to classical radiometry and geometrical optics.

\subsection{Classical Radiometry}

To relate the phase space light energy density $\ell \in \Den(T^*Q)$ to radiance, the central quantity in the classical formulation of radiative transfer, we have to consider measurements, the question Lambert was studying when he introduced the concept~\cite{Lambert2001}.
Measurements determine the flux of light energy density, for example through the sensor of a camera.
Mathematically, this flux can be determined using the transport theorem of tensor calculus~\cite[Theorem 8.1.12]{Marsden2004}. 
One then obtains that the energy $E$ flowing through a $2$-dimensional surface $M$ in a time interval $[t_1,t_2]$ is given by
\begin{align}
  E 
  = \int_{t_1}^{t_2} \! \! \int_{T^- M} i_{X_H} \ell
  \ = \int_{t_1}^{t_2} \! \! \int_{M} \frac{c}{n(q)} \int_{T_q^- M} \mathcal{L}(q,p) \, (\bar{p} \cdot \vec{n}(q)) \, dA \, d\bar{p} \, dt
  \label{eq:measurement}
\end{align}
where $\bar{p}$ is a unit vector, $\vec{n}(q)$ the surface normal of $M$ at $q$, and $T_q^{-} M$ the positive half-space of $T_q^* Q$ as defined by $\vec{n}(q)$~\cite[Chapter 3.2.6]{Lessig_phd}.
When the light energy density is parameterized in spherical coordinates, an infinitesimal measurement can be written as
\begin{align}
  \mathcal{L}(q,\bar{p},\nu) \, (\bar{p} \cdot \vec{n}) \, dA \, d\bar{p} 
  = \vec{n} \! \cdot \! \left( \mathcal{L}(q,\bar{p},\nu) \, \bar{p} \, dA \, d\bar{p} \, d\nu \right) 
\end{align}
and when no measurement surface, and hence no normal $\vec{n}$, is fixed one thus has for an infinitesimal measurement that
\begin{align}
  \Lambda = \mathcal{L}(q,\bar{p},\nu) \, dA_{\perp} \, d\bar{p} \, d\nu 
\end{align}
where $dA_{\perp}(\bar{p}) = \bar{p} \, dA$ is the standard area form for a surface orthogonal to the flow direction $\bar{p}$.
The differential $2$-form $\Lambda \in \Omega^2(Q)$ provides a modern interpretation of classical radiance.
The cosine term $(p \cdot \vec{n})$, which is prevalent in the classical literature but usually only justified heuristically~\cite{Nicodemus1963}, can then be obtained rigorously through the pullback of $\Lambda$ onto a surface with normal $\vec{n}$. 
We refer to~\cite{Lessig_phd} for the derivation of other concepts of classical radiometry such as vector irradiance.

\begin{remark}
Traditionally, it has often been overlooked that radiance is meaningful only in the context of measurements while the quantity that is naturally transported in radiative transfer is the phase space light energy density $\ell \in \Den(T^*Q)$.
This has led to considerable confusion even in recent literature~\cite{Bal2006}. 
\end{remark}

\subsection{Radiative Transfer and Geometrical Optics}

\begin{figure}[t]
  \setlength{\abovecaptionskip}{-10pt}
  \begin{center}
    \centerline{\hspace{0.0in}
    \includegraphics[trim = 30mm 200mm 90mm 37mm, clip, scale=1.0]{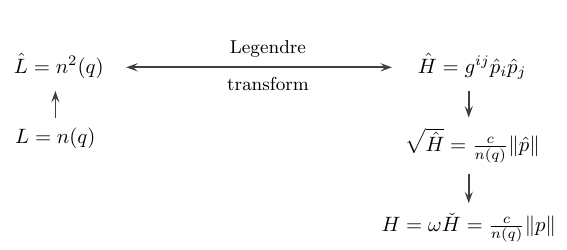}}
  \end{center}
  \caption{Non-canonical Legendre transform relating radiative transfer to geometrical optics given by Fermat's principle.}
  \label{fig:legendre}
\end{figure}

A question rarely considered in the classical literature on radiative transfer is the relationship of the theory to geometrical optics. 
The connection can be established by considering the Legendre transform of Fermat's principle.
As is well known, directly performing the transform for the Lagrangian $L = n(q)$ leads to a vanishing Hamiltonian~\cite{Kline1965}. 
Following Arnold~\cite{Arnold1989},\footnote{We were told this idea goes back at least to Riemann.} instead of length, given by $L = n(q)$, we shall hence consider the geometrical energy of a light path, given by $\hat{L} = n^2(q)$.
This Lagrangian can be interpreted as a diagonal metric $g_{ij} = n^{2}(q) / c^2 \, \delta_{ij}$ and the associated geodesic flow is equivalently described by the Hamiltonian $\hat{H} = g^{ij} \, \hat{p}_i \, \hat{p}_j$~\cite[p. 51]{Jost2008} where $\hat{p}$ is the canonical momentum which is related to the kinetic momentum by $p = \omega \, \hat{p}$.
Reverting the transition from path length to path energy, and including the factor of $\omega$ corresponding to energy we obtain for the Hamiltonian again Eq.~\ref{eq:hamiltonian}.
If we trace the diagram in Fig.~\ref{fig:legendre} backwards, we see that, from the point of view of radiative transfer, geometrical optics is a limiting case when the amount of energy that is transported is disregarded.

%% file: conclusion.tex
\section{Discussion and Open Questions}
\label{sec:conclusion}

Our geometric formulation of radiative transfer and the identification of the Lie group structures that underlie the known conservation laws clarifies and unifies earlier work in the literature.
Additionally, the use of tensor calculus overcomes the limitations of the current formulation, for example when measurements are considered, and it improves over earlier attempts that employed vector calculus~\cite{Gershun1936} and measure theory~\cite{Preisendorfer1965} to obtain a modern mathematical foundation for radiative transfer.\footnote{The status and shortcomings of many classical derivations of radiative transfer theory were recently summarized by Mishchenko~\cite{Mishchenko2011a}.}
The derivation of radiative transfer from electromagnetic theory that was  presented in Sec.~\ref{sec:derivation} largely follows recent work in applied mathematics~\cite{Gerard1997,Ryzhik1996}, which can be seen as a refinement of earlier but little known results in plasma physics.\footnote{See~\cite{Pomraning1973} and references therein and~\cite{Wolf1976}. In theoretical optics, various alternative names are employed for the Wigner transform, cf.~\cite{Bastiaans1986}.}
Our presentation emphasized geometric aspects of the argument and it completed the connection to the classical formulations in the literature~\cite{Gerard1997,Ryzhik1996,Bal2006}.
Nonetheless, the structures that underlie many aspects of the derivation remain currently unclear.
In the following, we will collect some preliminary results on how to fill these gaps.

\begin{figure*}[t]
  \setlength{\abovecaptionskip}{-15pt}
  \setlength{\belowcaptionskip}{-5pt}
  \begin{center}
  \centerline{
  \includegraphics[trim = 48mm 75mm 60mm 62mm, clip, scale=0.6]{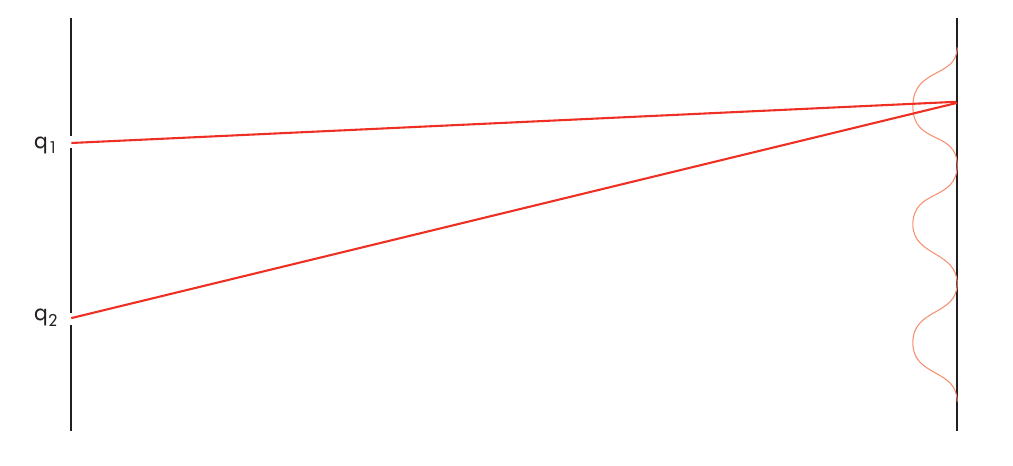}}
  \end{center}
  \caption{Young's experiment: coherent electromagnetic radiation passes through the double slit formed by $q_1$ and $q_2$ and through interference forms the intensity pattern on the screen on the right.}
  \label{fig:youngs_experiment}
\end{figure*}

Additional insight into the derivation in Sec.~\ref{sec:derivation} can be obtained by considering a density matrix formulation of electromagnetic theory before the phase space lift.
In quantum mechanics, the density matrix for a pure state $\psi$ is defined by $\rho = \psi \rangle \langle \psi$ and it represents the projection operator onto the one dimensional subspace spanned by $\psi$.\footnote{The density matrix was introduced in a famous paper by von Neumann~\cite{vonNeumann1927} to study statistical ensembles of states, an aspect we will not consider here but which is closely related to the questions considered in statistical optics, cf.~\cite{Mandel1995}.}
One of the advantages of this formulation is that it provides a faithful representation of the projective Hilbert space $\mathbb{CP}^n$ that serves as the configuration space of quantum mechanics, cf.~\cite[Chapter 5.4.3]{Ratiu1999}.\footnote{This representation of $\mathbb{CP}^n$ is the prototypical example of a $C^*$- or von Neumann algebra.}
The density matrix for the electromagnetic field is known as the mutual coherence matrix~\cite{Mandel1995,Wolf2007} and there typically defined as\footnote{In the statistical optics literature one typically considers statistical averages of the field components, which we omit here. This is the analogue of the probabilistic superposition of pure states in quantum mechanics.}
\begin{align}
  P_{ij}(q,t,\bar{q},\bar{t}) = F_i(q,t) \, F_j^*(\bar{q},\bar{t}) .
  \label{eq:density_matrix}
\end{align}
Analogous to the situation in quantum mechanics, the trace $\tr{(P)}$ of the density matrix, for $(q,t) = (\bar{q},t)$, is proportional to the quadratic observable, the electromagnetic energy density $\mathcal{E}(q,t)$~\cite{Roychowdhury2003}. 
Neither Eq.~\ref{eq:density_matrix} nor the trace have an apparent geometric interpretation. 
However, we know from Eq.~\ref{eq:em_energy_density} that the energy density is given by $\mathcal{E} = \Vert \vec{F} \Vert_{\eps,\mu}^2$. 
Using the Faraday $2$-form $F = E \wedge dt + B$ this can be written as 
\begin{align}
    \mathcal{E}(q,t) &= \langle \! \langle F , F \rangle \! \rangle_{\eps,\mu} = F \wedge \star_{\eps,\mu} F
    \label{eq:em_energy_density:wedge_product:1}
\end{align}
where $\star_{\eps,\mu}$ is the Hodge dual induced by considering the electric permittivity and magnetic permeability as part of the metric. By definition of the wedge product, this is equivalent to
\begin{align}
    \mathcal{E}(q,t) &= \mathrm{A}( F \otimes \star_{\eps,\mu} F )
    \label{eq:em_energy_density:wedge_product:2}
  \end{align}
where $\textrm{A}$ is the anti-symmetrization map~\cite[Def. 7.1.3]{Marsden2004}. 
As can be shown by a straightforward computation, the anti-symmetrization in Eq.~\ref{eq:em_energy_density:wedge_product:2} is, in flat spacetime, equivalent to taking the trace of $\vec{F} \otimes \vec{F}$. 
We hence indeed have
\begin{align}
  \mathcal{E}(q,t) = \mathrm{A}( F \otimes \star_{\eps,\mu} F ) = \tr{(\vec{F} \otimes \vec{F})} .
  \label{eq:em_energy_density:density_matrix}
\end{align}
We believe that $F \otimes \star_{\eps,\mu} F$ provides a mathematically and physically more natural definition of the density matrix. 
The non-locality in the definition in Eq.~\ref{eq:density_matrix} can be understood by considering interference phenomena such as those arising in the classical Young's interference experiment, cf. Fig.~\ref{fig:youngs_experiment}.
There, interference arises from the superposition of the fields at the pinholes, and the interference fringes, and hence the intensity of the electromagnetic field, can be described through the nonlocal coherence matrix $P(q,t,\bar{q},t)$, see~\cite{Wolf2007}. We believe that this idea can be made rigorous by considering the time dependence for $F$ in Eq.~\ref{eq:em_energy_density:density_matrix} and exploiting that $F_t = U_t F_0$ where $U_t$ is a unitary operator. 

As in the case of the Schr{\"o}dinger equation, differentiating the definition of the local density matrix with respect to time using the Leibniz rule and inserting Eq.~\ref{eq:maxwell:operator} in the resulting expression yields $\dot{P} = - [ P , M ]$ where $[\, , ]$ denotes the matrix commutator. 
As is well known~\cite{Evans2011,Wunsch2008}, under the semiclassical symbol calculus the commutator becomes the Moyal bracket on phase space, and, at least formally, it can be shown that the Wigner transform is the symbol of the density operator.
Using the density matrix and its time evolution equation provides thus a more natural transition from Maxwell's equations on configuration space to Moyal bracket dynamics on phase space.

For the Schr{\"o}dinger equation, the model problem in semiclassical analysis, the representation theory of the Heisenberg group plays a central role, as is evident from the Stone-von Neumann theorem, which, roughly speaking, states that all formulations of quantum mechanics are essentially unitarily equivalent. 
From the point of view of the Heisenberg group, the short wavelength limit is the group contraction that yields the symplectic group.
A question of interest to us is to understand which role the Heisenberg group plays for the asymptotic limit discussed in this paper.
Interesting work in this context is for example those by Landsman~\cite{Landsman1998} who discusses connections between Lie-Poisson reduction and quantization using the Heisenberg group. 

Quite curious in the derivation in Sec.~\ref{sec:derivation} are the matrix-valued ``Moyal'' and ``Poisson'' brackets that arise for example in Eq.~\ref{eq:moyal_bracket:expanded} and Eq.~\ref{eq:transport:polarization}. 
Although these brackets are known in the physics literature~\cite[Chapter 16.3]{Spohn2004}~\cite[Appendix A]{Teufel2003}, and they also appear in the microlocal and semiclassical analysis literature, cf.~\cite{Gerard1997}, they have to our knowledge not been studied from a geometric point of view. 
One approach to generalize the Poisson bracket to the matrix-valued case is to consider
\begin{align}
  \dot{f} = - \{ f , H \} = - \Lie_{X_H} f
\end{align}
in which case the right hand side has a natural extension for arbitrary tensors.
For a matrix $A$, that is a $(1,1)$ tensor, one then obtains
\begin{align}
  \dot{A} = - \Lie_{X_H} A = - \{ A , H \} + [ A , \bar{H} ]
  \label{eq:lie_derivative:matrix}
\end{align}
where the ``Poisson bracket'' for the matrix $A$ is defined as before, as a component wise bracket, and $\bar{H}$ is the Hessian ``matrix'' of the Hamiltonian, that is the matrix of second partial derivatives.
Eq.~\ref{eq:lie_derivative:matrix} has the same form as Eq.~\ref{eq:transport:polarization} although it is currently not clear to us under which conditions $\bar{H}$ coincides with the first order term $m^1$ of the symbol.
For the situation where also the generator of the dynamics is matrix-valued, the connection between the symbol of an operator and the dispersion matrix, which is well understood from a physical point of view, seems to play a key role, cf.~\cite{Ryzhik1996}.
Preliminary work in the literature that considers matrix-valued quantization from a geometric perspective is~\cite{Littlejohn1991a,Emmrich1996,Bursztyn2000}, although to our knowledge no complete picture exists at the moment.

An interesting open question is also the transition from Maxwell's equations to radiative transfer in spacetime, the natural setting of electromagnetic theory. 
Although no general theory of covariant Poisson brackets in spacetime exists, for the special case of Maxwell's equations a bracket is known~\cite{Marsden1986a}.
Moreover, the Faraday $2$-form plays from the outset an important role in our derivation and semiclassical analysis naturally considers spacetime operators, cf.~\cite{Wunsch2008}.
A derivation in this setting might also help to understand how the structure of electromagnetic theory manifests itself at the short wavelength limit and how the symmetries of radiative transfer theory arise.

Despite many connections, microlocal and semiclassical analysis are currently rarely considered in geometric mechanics.
We believe this is an area ripe for further investigations. 
For example, in many situations microlocal analysis also allows the description of a Hamiltonian system on phase space $T^*Q$ through an equivalent partial differential equation on configuration space $Q$. 
We believe that this provides additional insight into the plasma-to-fluid map~\cite{Marsden1982a} and might allow to generalize the result.
It would also be interesting to explore how existing results, for example for the Maxwell-Vlasov system~\cite{Marsden1982a} or Euler-Yang-Mills fluids, cf.~\cite{Gay-Balmaz2008}, can be reformulated when electromagnetic theory is describes on phase space. 

Although the Hamiltonian formulation of radiative transfer has been known in plasma physics for a long time~\cite{Pomraning1973}, it has so far not been appreciated in other communities. 
We belief that the $5$-dimensional formulation of radiative transfer that is prevalent in the literature, and which is incompatible with a Hamiltonian description that necessitates an even dimensional phase space, led to much confusion on the subject. 
Our reduction of the $6$-dimensional Hamiltonian system to a contact Hamiltonian system on the cosphere bundle $S^*Q$ clarifies this relationship.
The Lie-Poisson structure of radiative transfer mirrors those of other systems in statistical mechanics whose time evolution is describes by the Vlasov equation~\cite{Marsden1982a,Tronci_phd}. 
Nonetheless, since radiative transfer is rarely written in the form of Eq.~\ref{eq:lte} it was surprising to us that the classical law of conservation of radiance arises from a Lie-Poisson structure. 
Similarly, the structural similarities between ideal fluid dynamics and ideal light transport in Table~\ref{tab:comp:fluid_light} seem, from the point of view of the classical literature, quite remarkable.


\section{Conclusion}

This work owes much to Jerry Marsden, to his encouragement, and to his writings.
Jerry told us that if there is a geometric formulation of radiative transfer then it is worth developing it. 
We always reminded ourselves of this when nothing seemed to fit together.
Jerry's writings also repeatedly provided us a life line and they made geometric mechanics accessible to us.